\documentclass[12pt,a4paper]{article}
\pdfoutput=1
\usepackage{jcappub}
\usepackage{amsmath}
\usepackage{latexsym}
\usepackage{amsfonts}
\usepackage{mathrsfs}

\title{Primordial black holes and second order gravitational waves from ultra-slow-roll inflation}

\author{Haoran Di}
\author[1]{and Yungui Gong\note{Corresponding author}}
\affiliation{School of Physics, Huazhong University of Science and Technology,
Wuhan, Hubei 430074, China}

\emailAdd{hrd@hust.edu.cn}
\emailAdd{yggong@hust.edu.cn}

\abstract{
The next generation of space-borne gravitational wave detectors may detect gravitational waves from extreme
mass-ratio inspirals with primordial black holes.
To produce primordial black holes which contribute a non-negligible abundance of dark matter and are consistent
with the observations, a large
enhancement in the primordial curvature power spectrum is needed. For a single field slow-roll inflation,
the enhancement requires a very flat potential for the inflaton, and this will increase the number of $e$-folds.
To avoid the problem, an ultra-slow-roll inflation at the near inflection point is required. We elaborate
the conditions to successfully produce primordial black hole dark matter from single field inflation and propose a toy model
with polynomial potential to realize the big enhancement of the curvature power spectrum at small scales
while maintaining the consistency with the observations at large scales. The power spectrum for the second order
gravitational waves generated by the large density perturbations at
small scales is consistent with the current pulsar timing array observations.}

\begin{document}
\maketitle

\section{Introduction}

The detection of the gravitational waves (GWs) from black hole mergers by the LIGO Collaboration
and Virgo Collaboration opens a new window to probe black hole physics \cite{Abbott:2016blz,Abbott:2016nmj,Abbott:2017vtc,Abbott:2017gyy,Abbott:2017oio}.
The next generation of space-borne GW detectors will operate in the $0.1-100$mHz frequency band
and detect GWs from supermassive black hole binaries, Galactic white-dwarf binaries and extreme
mass-ratio inspirals with primordial black holes (PBHs) \cite{Hawking:1971ei,Carr:1975qj}.
PBH can be taken as dark matter candidate. Due to the failure of direct detection of particle dark matter
such as weakly interacting massive particles and axions, the interest in PBH as dark matter has grown recently
\cite{Khlopov:2004sc,Belotsky:2014kca,Frampton:2010sw,Drees:2011hb,Jacobs:2014yca,Kawasaki:2016pql,Harada:2013epa,
Clesse:2015wea,Carr:2016drx,Bird:2016dcv,Clesse:2016vqa,Inomata:2017okj,Garcia-Bellido:2017mdw,
Garcia-Bellido:2017fdg,Kovetz:2017rvv,Motohashi:2017kbs,Kuhnel:2017bvu,Germani:2017bcs,Carr:2017jsz}.
PBHs with mass smaller than $10^{15}$g would have evaporated by now through Hawking radiation \cite{Carr:1975qj}.
Observations from femtolensing of gamma-ray bursts \cite{Gould:1992apjl,Nemiroff:2001bp}, millilensing of compact radio sources \cite{Wilkinson:2001vv},
microlensing of quasars \cite{Dalcanton:1994apj}, the Milky way and Magellanic Cloud
stars \cite{Allsman:2000kg,Tisserand:2006zx,Carr:2009jm,Griest:2013esa}
constrained the mass range of PBHs to be from $10^{15}$g to $2\times 10^{17}$g and from $2\times 10^{20}$g
to $4\times 10^{24}$g \cite{Jacobs:2014yca}.
However, it was shown that the mass window for PBH as all dark matter is around $10^{20}$g \cite{Inomata:2017okj},
or around $5\times 10^{-16}M_\odot$, $2\times 10^{-14}M_\odot$ and $25-100M_\odot$ \cite{Carr:2017jsz}, so the mass window for PBH is narrow
and needs further study.

PBH forms in the radiation era as a result of gravitational collapse of density perturbations generated during inflation.
To produce appreciable abundance of PBH dark matter, the curvature power spectrum needs to be amplified to the order of 0.01
near the end of inflation, so the slow-roll parameter $\epsilon$ is required to decrease by at least 7 orders of magnitude.
This may be realized at a near inflection point where the potential becomes almost a constant \cite{Garcia-Bellido:2017mdw,Ezquiaga:2017fvi,Cicoli:2018asa}.
The models introduced in \cite{Garcia-Bellido:2017mdw,Ezquiaga:2017fvi}
avoid the violation of slow-roll and the jump in $N$ at the near inflection point is around $\Delta N\sim 30$, but
the enhancement of the power spectrum is less than 5 orders of magnitude.
For typical potentials with a inflection point, the problem is that inflation either ends or the slow-roll
condition is violated before reaching the near inflection point\cite{Kannike:2017bxn,Germani:2017bcs,Motohashi:2017kbs}.
On the other hand, the deep decrease in $\epsilon$ will increase the number of $e$-folds $N$.
Furthermore, it is even possible that the sudden change in $\epsilon$ lasts for as long as 60 $e$-folds, then the power
spectrum is featureless only in a narrow range of scales around $k_*$, and
the $\mu$ distortion of the power spectrum becomes large.
Therefore, to get the large enhancement of the curvature power spectrum from the single field slow-roll inflation,
we expect the following problems:
(1) the slow-roll approximation breaks down around the near inflection point; (2) the number of $e$-folds $N$ is much larger than $60$;
(3) the big change in $\epsilon$ lasts over 60 $e$-folds and the $\mu$ distortion is large; (4)
the large enhancement of the power spectrum may induce large second order gravitational wave signals \cite{Matarrese:1997ay,Mollerach:2003nq,Ananda:2006af,Baumann:2007zm,Garcia-Bellido:2017aan,
Saito:2008jc,Saito:2009jt,Bugaev:2009zh,Bugaev:2010bb,Alabidi:2012ex,Orlofsky:2016vbd,Nakama:2016gzw,Inomata:2016rbd,Cheng:2018yyr}.

For the slow-roll inflation, naively we expect that the inflaton almost stop rolling when the potential becomes very flat,
and the number of $e$-folds becomes very large.
However, if the potential is very flat, then the inflaton enters the ultra-slow-roll inflation \cite{Tsamis:2003px,Kinney:2005vj}
and the analytical expressions for the power spectra from slow-roll inflation cannot be applied because one of the slow-roll conditions is violated \cite{Yi:2017mxs}.
For the ultra-slow-roll inflation, the acceleration of the inflaton is locked by the friction term
and the inflaton continues to roll down the potential \cite{Motohashi:2014ppa,Dimopoulos:2017xox,Dimopoulos:2017ged}, so the number of $e$-folds
spent at the near inflection point becomes smaller than expected.
Therefore, a period of ultra-slow-roll inflation helps alleviate the large $N$ problem.
In this paper, we give the conditions for the
reach of the ultra-slow-roll inflation and use a toy model with the polynomial potential to show that the successful production
of the primordial dark matter can be achieved by ultra-slow-roll inflation and the problems for the slow-roll inflation do not exist
for the ultra-slow-roll inflation.

\section{Primordial black holes from inflation}

By using the Press-Schechter theory and smoothing on the scale $M$ with a Gaussian,
the fractional energy density of PBH in the Universe is
\begin{equation}
\label{omegabh}
\beta=\frac{\rho_{\text{PBH}}}{\rho_{\text{tot}}}
\approx \text{erfc}\left(\frac{\delta_c}{\sqrt{2\mathcal{P}_\delta}}\right)=\text{erfc}\left(\frac{9\delta_c}{4\sqrt{2\mathcal{P}_\zeta}}\right),
\end{equation}
where the critical density perturbation is assumed to be in the range $\delta_c=0.07-0.7$ \cite{Carr:1975qj,Niemeyer:1997mt,Niemeyer:1999ak,Green:2004wb,Musco:2004ak,Harada:2013epa,Clesse:2015wea,Carr:2016drx},
the density contrast $\delta$ is related with the primordial curvature perturbation $\zeta$ as $\delta=4\zeta/9$ during radiation domination,
and the power spectrum for the primordial scalar perturbation is
\begin{equation}
\label{pwrspec1}
\mathcal{P}_\zeta=\frac{H^2}{8\pi^2\epsilon}\approx \mathcal{P}_\zeta(k_*)\frac{V(\phi)}{V(\phi_*)}\frac{\epsilon(\phi_*)}{\epsilon(\phi)},
\end{equation}
where we choose $M_{\text{Pl}}^{-2}=8\pi G=1$,
the slow-roll parameter $\epsilon=\frac{1}{2}(V_\phi/V)^2$, $V_\phi=dV/d\phi$,
and the $\phi_*$ is the value of $\phi$ at the time when a given scale $k_*$ exits the horizon.
In this paper, we choose $k_*=0.05$Mpc$^{-1}$.
Note that the formula \eqref{pwrspec1} is valid only with the slow-roll approximation.
PBH with mass greater than $M$ forms when the density contrast $\delta$ exceeds the critical value $\delta_c$.
The mass $M$ of the PBH that forms during radiation domination is of the same order of the horizon mass $M_H$,
$M=\gamma M_H$, where
\begin{equation}
\label{mheq1}
M_H=\frac{4\pi\rho}{3H^3}=\frac{4\pi}{H}=\Omega_r^{1/2}M_0\left(\frac{g}{g_0}\right)^{-1/6}\left.\left(\frac{H_0}{k}\right)^2\right|_{k=aH},
\end{equation}
we neglect the difference between the effective degrees of freedom in the entropy and energy densities and denote them as $g$,
the current value of $g$ is $g_0=3.36$, $g=107.5$ for $T>300$GeV and $g=10.75$ for $0.5\text{MeV}<T<300\text{GeV}$,
$\Omega_r$ is the radiation energy density today,
the Hubble constant $H_0=100h\ \text{km/s/Mpc}$, and
$M_0=4.63\times 10^{22}M_\odot(0.6727/h)$. We choose the current temperature for the cosmic microwave background radiation $T_0=2.725$ K, so $\Omega_r h^2=4.15\times 10^{-5}$.
Since its formation, the energy density $\rho_{\text{PBH}}$ of PBH scales slower than the total energy density $\rho_{\text{tot}}$ of the Universe during radiation domination,
so the relative contribution to the total energy density grows. Ignoring the mass accretion and evaporation and assuming that $\beta(M)$
is a constant, we have \cite{Carr:2016drx,Inomata:2016rbd,Inomata:2017okj,Motohashi:2017kbs,Garcia-Bellido:2017aan,Nakama:2016gzw,Orlofsky:2016vbd}
\begin{equation}
\label{omegabh2}
\beta=1.76\times 10^{-9}\gamma^{-1/2}\left(\frac{\Omega_{\text{PBH}}h^2}{0.12}\right)\left(\frac{g}{10.75}\right)^{1/4}\left(\frac{M}{M_\odot}\right)^{1/2},
\end{equation}
and the current fractional energy density of the PBHs to dark matter
\begin{equation}
\label{fomh2}
f_{\text{PBH}}=\frac{\Omega_{\text{PBH}}h^2}{\Omega_c h^2}=5.68\times 10^8\beta(M)\gamma^{1/2}\frac{0.12}{\Omega_c h^2}\left(\frac{g}{10.75}\right)^{-1/4}\left(\frac{M}{M_\odot}\right)^{-1/2},
\end{equation}
where $\Omega_c$ is the current energy density of dark matter.
If all the dark matters are PBHs with $M=(10^{-13}M_\odot, 2\times 10^{-9}M_\odot)$ \cite{Jacobs:2014yca,Kuhnel:2017bvu} and $\gamma=1$,
then we get $\beta=(9.76\times 10^{-16}, 1.38\times 10^{-13})$. If $M=(5\times 10^{-19}M_\odot, 10^{-16}M_\odot)$ \cite{Jacobs:2014yca,Kuhnel:2017bvu}, here the lower
bound is set by Hawking radiation \cite{Carr:1975qj}, then $\beta=(2.2\times 10^{-18}, 3.1\times 10^{-17})$.

Since $\mathcal{P}_\zeta(k_*)=2.2\times 10^{-9}$ \cite{Ade:2015lrj}, for appreciable dark matter to be in the form of PBHs, we need a large enhancement on $\mathcal{P}_\zeta$,
for example, at least 7 orders of enhancement from the slow-roll parameter $\epsilon$.
This requires the potential to be very flat at the enhancement point.
To keep inflation, the enhancement point should be a near inflection point. However,
this enhancement not only decreases the field excursion \cite{Gao:2014pca}, but also increases the number of $e$-folds because
\begin{equation}
\label{nefoldeq1}
N=\int_\phi^{\phi_e} \frac{d\phi}{\sqrt{2\epsilon(\phi)}}.
\end{equation}
At the near inflection point $\phi_{\text{infl}}$, the slow-roll parameter decreases by 7 orders of magnitude,
unless this change happens around $\Delta\phi=10^{-6}$ or less,
it will contribute a lot to $N$. In fact, almost all the number of $e$-folds comes from the contribution around $\phi_{\text{infl}}$,
then the power spectrum gets big boost in the small scales, this will
cause big $\mu$ distortion \cite{Inomata:2016rbd,Nakama:2017ohe}
because the $\mu$ distortion is \cite{Chluba:2015bqa,Nakama:2017ohe}
\begin{equation}
\label{mueq1}
\mu_{\text{ac}}\approx \int_{k_{\text{min}}}^\infty \frac{dk}{k}\mathcal{P}_\zeta(k)W_\mu(k),
\end{equation}
where
\begin{equation}
\label{mueq2}
W_\mu(k)=2.8A^2\left[\exp\left(-\frac{[\hat{k}/1360]^2}{1+[\hat{k}/260]^{0.3}+\hat{k}/340}\right)
-\exp\left(-\left[\frac{\hat{k}}{32}\right]^2\right)\right],
\end{equation}
$k_{\text{min}}\approx 1$Mpc$^{-1}$, $A\approx 0.9$ and $\hat{k}=k/[1\ \text{Mpc}^{-1}]$.

If $\epsilon$ changes too quickly
at the near inflection point, then the slow-roll condition will be violated \cite{Motohashi:2017kbs} because
\begin{equation}
\label{depsiloneq1}
\left|\frac{d\epsilon}{d\phi}\right|=\sqrt{2\epsilon}|\eta-2\epsilon|,
\end{equation}
where $\eta=V_{\phi\phi}/V$. So either $\epsilon$ or $\eta$ will be larger 1 if $|d\epsilon/d\phi|\gg 1$.
Therefore, we expect either of the following problems
for single field inflation: (1) the slow-roll approximation breaks down around the
near inflection point $\phi_{\text{infl}}$; (2) $N$ is much larger than $60$;
(3) the power spectrum is enhanced over almost the whole range of the scales $k>k_*$ and the $\mu$ distortion becomes large.

However, if an ultra-slow-roll inflation in reached,
then the slow-roll formula \eqref{nefoldeq1} may overestimate the number of $e$-folds
because the acceleration of the inflaton is locked by the friction term \cite{Dimopoulos:2017xox,Dimopoulos:2017ged}.
For example, in the critical Higgs inflation with the potential $V(\phi)=\lambda(\phi)\phi^4/4$
and the nonminimal coupling $\xi(\phi)\phi^2 R$ \cite{Ezquiaga:2017fvi}, where
\begin{gather}
\label{eq11}
\lambda(\phi)=\lambda_0+b_\lambda\ln^2(\phi/\mu),\\
\label{eq12}
\xi(\phi)=\xi_0+b_\xi\ln(\phi/\mu),
\end{gather}
an ultra-slow-roll inflation is reached around the near inflection point $\phi/\mu=0.785$ and we get the enhancement on
the power spectrum by about three orders of magnitude
if we take $\lambda_0=2.69\times 10^{-7}$, $\xi_0=9.22$, $\mu^2=0.118$, $b_\lambda=1.1\times 10^{-6}$ and $b_\xi=10.9$ \cite{Ezquiaga:2017fvi}.
In this model, the number of $e$-folds spent on the near inflection point is only $\Delta N=35$.
However, the enhancement is not big enough to produce significant PBH dark matter.

\begin{figure}
\centerline{$\begin{array}{cc}
\includegraphics[width=0.35\textwidth]{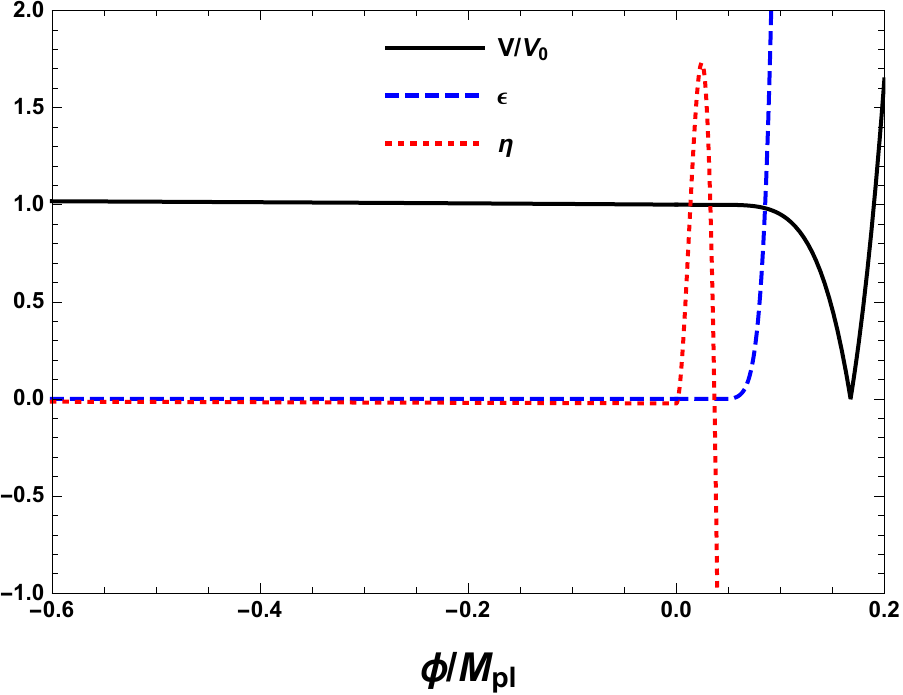}\quad \quad &
\includegraphics[width=0.35\textwidth]{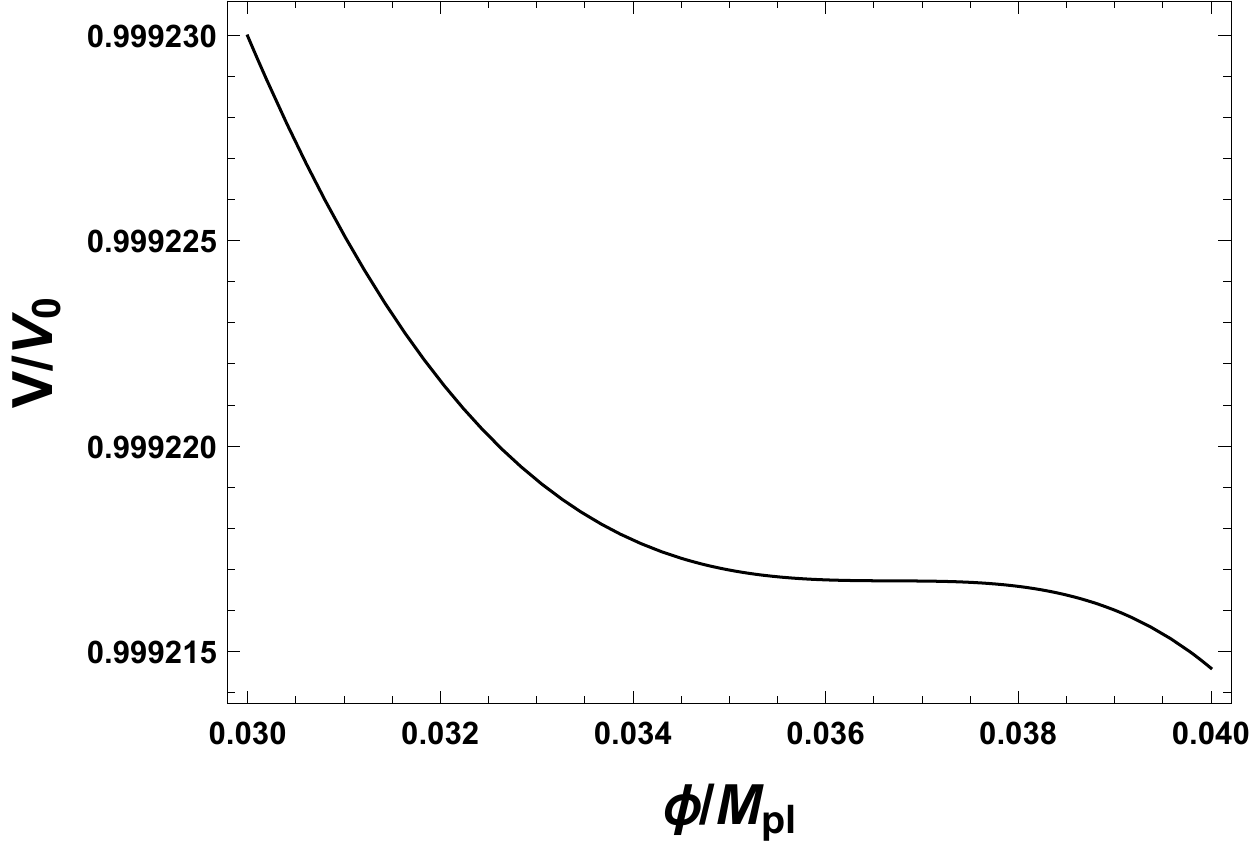}
\end{array}$}
\caption{The potential along with its slow-roll parameters for the polynomial model.
The right panel shows the behaviour of the potential around the inflection point.}
\label{fig1}
\end{figure}

We propose a toy model with the polynomial potential to produce a significant amount of PBH dark matter without the problems for single slow-roll inflation mentioned above.
We consider the polynomial potential
\begin{equation}
\label{vphieq1}
V(\phi)=\begin{cases}
\displaystyle{V_0\left|1+\sum_{m=1}^{m=5}\lambda_m\left(\frac{\phi}{M_{\text{Pl}}}\right)^m\right|},& \phi\ge 0,\\
\displaystyle{V_0\left[1+\sum_{m=1}^{m=3}\lambda_m\left(\frac{\phi}{M_{\text{Pl}}}\right)^m\right]},& \phi<0,
\end{cases}
\end{equation}
where the first three coefficients $\lambda_1$, $\lambda_2$ and $\lambda_3$ are determined from $n_s$, $r$ and $n_s'=dn_s/d\ln k$,
the coefficients $\lambda_4$ and $\lambda_5$ are determined from the near-inflection
condition $V_\phi\approx 0$ and $V_{\phi\phi}\approx 0$ \footnote{This is the reason why we need to consider a polynomial with at least fifth degree.}.
The polynomial potential may be obtained from supergravity model building \cite{Gao:2014pca}.
Choosing $\phi_*=-0.54$ and $\phi_{\text{infl}}=0.0367$, and using the Planck 2015 results, $k_*=0.05$Mpc$^{-1}$, $n_s=0.9686$, $r=0.005$,
$n_s'=-0.0008$ and $\mathcal{P}_\zeta=2.2\times 10^{-9}$ \cite{Ade:2015lrj}, we get
$\lambda_1=-0.0353553$, $\lambda_2=-0.0115783$, $\lambda_3=-0.00235702$, $\lambda_4=728.239$, $\lambda_5=-11882.9$
and $V_0=1.55\times 10^{-10}$. The potential along with its slow-roll parameters $\epsilon$ and $\eta$ are shown in Fig. \ref{fig1}.
From Fig. \ref{fig1}, we see that the slow-roll parameter $\eta>1$ before the inflaton reaches the near inflection point $\phi_{\text{infl}}$.
Even though $\eta$ becomes larger than 1 briefly, the slow-roll parameter $\epsilon\ll 1$ and the inflation does not end.
Furthermore, this property is important for the reach of the ultra-slow-roll inflation.

\begin{figure}
\begin{center}
$\begin{array}{cc}
\includegraphics[width=0.35\textwidth]{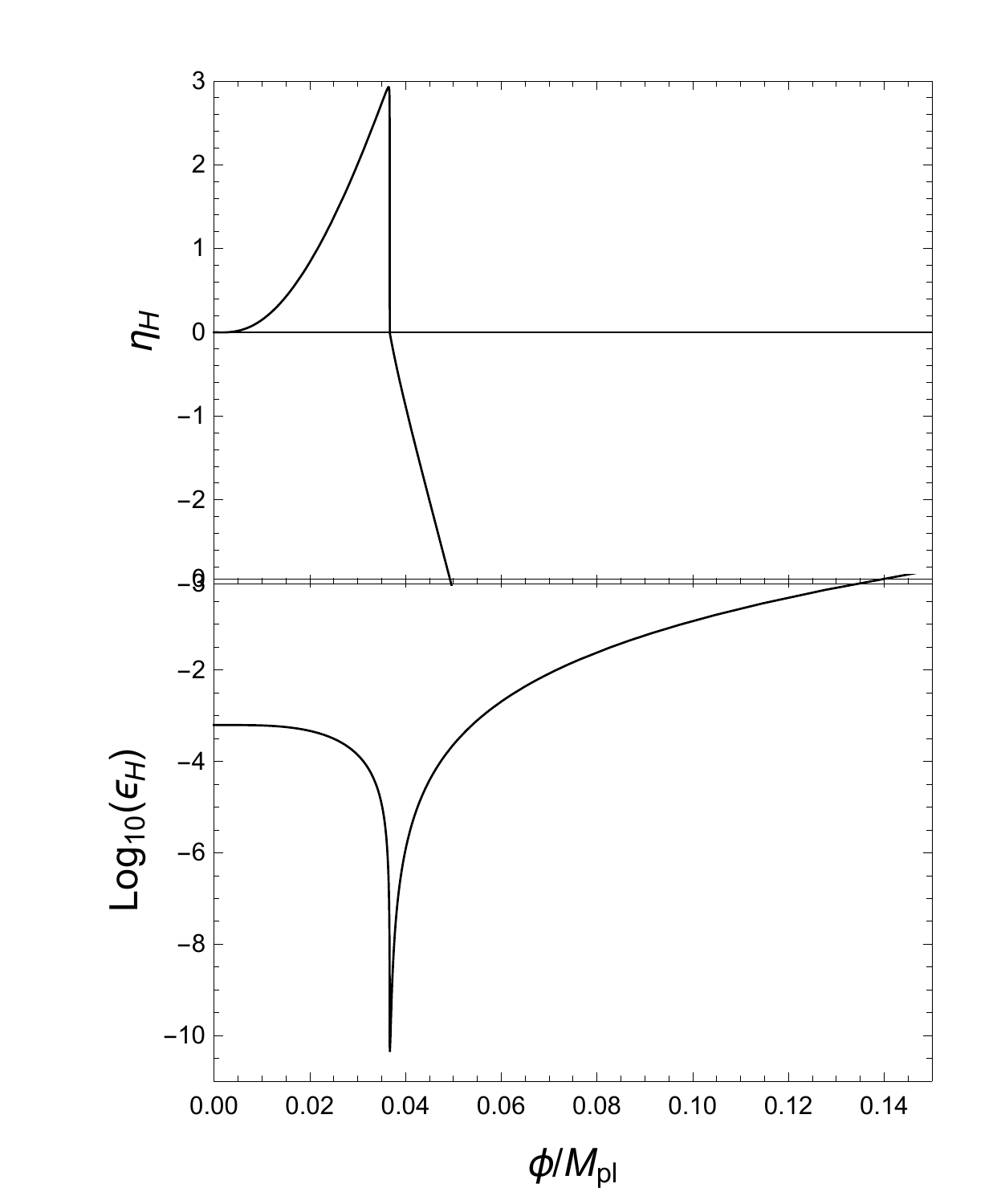}\quad \quad &
\includegraphics[width=0.35\textwidth]{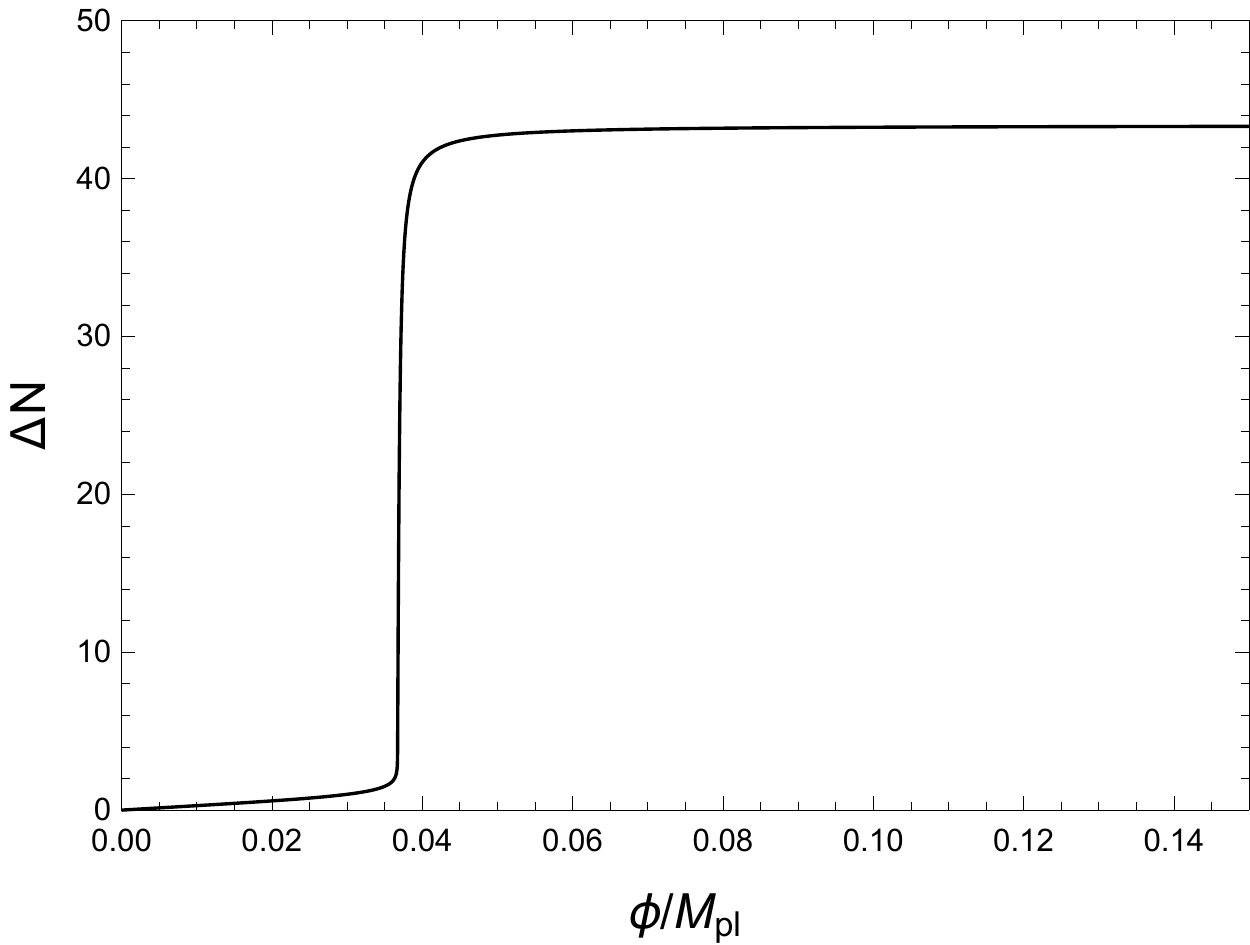}
\end{array}$
\end{center}
\caption{The left panel shows the evolutions of the Hubble flow slow-roll parameter $\epsilon_H$ and $\eta_H$.
The right panel shows the evolutions of the number of $e$-folds $\Delta N=\ln[a(\phi)/a(0)]$.}
\label{fig1a}
\end{figure}

Since the slow-roll parameter $\eta_H\approx 3$ near the inflection point, so one of the slow-roll
conditions is violated and the formula \eqref{pwrspec1} cannot be used to calculate the power spectrum near the inflection point.
By solving the equation of motion numerically,
we find that inflation ends at $\phi_e=0.14$, the inflaton excursion is $\Delta\phi=\phi_e-\phi_*=0.68 M_{\text{Pl}}$,
and the number of $e$-folds $N$ before the end of
inflation when the scale $k_*=0.05$Mpc$^{-1}$ exits the horizon
is $N=62.6$.
We show the evolution of the scale factor $a(t)$, the Hubble flow slow-roll parameter
$\epsilon_H=-\dot{H}/H^2$ and $\eta_H=-\ddot{\phi}/(H\dot{\phi})$ in Fig. \ref{fig1a}.
From Fig. \ref{fig1a}, we see that we have the ultra-slow-roll inflation
with $\eta_H\approx 3$ around $\phi_{\text{infl}}$,
the number of $e$-folds spent at $\phi_{\text{infl}}$ is $\Delta N=42$
and $\epsilon_H$ decreases by 7 orders of magnitude at $\phi_{\text{infl}}$.
Note that inflation is kept because $\epsilon_H\ll 1$ even though $\eta>1$ around $\phi_{\text{infl}}$.
The condition $\eta>1$ before $\phi_{\text{infl}}$ guarantees the reach of ultra-slow-roll inflation at $\phi_{\text{infl}}$ because
the condition causes $V'$ to decrease faster so that it becomes much smaller than the friction term $3H\dot{\phi}$ at $\phi_{\text{infl}}$.
To ensure enough number of $e$-folds and keep the power spectrum to be smooth and featureless over a wide range of scales,
we also require a long period of slow-roll inflation, this can be achieved by choosing the potential to be a cubic polynomial.
This is the reason we take the cubic polynomial for $\phi<0$. The long period of slow-roll inflation at large scales also helps to
avoid the initial value problems for the equation of motions because the slow-roll inflation is an attractor.

\begin{figure}
\centerline{\includegraphics[width=0.6\textwidth]{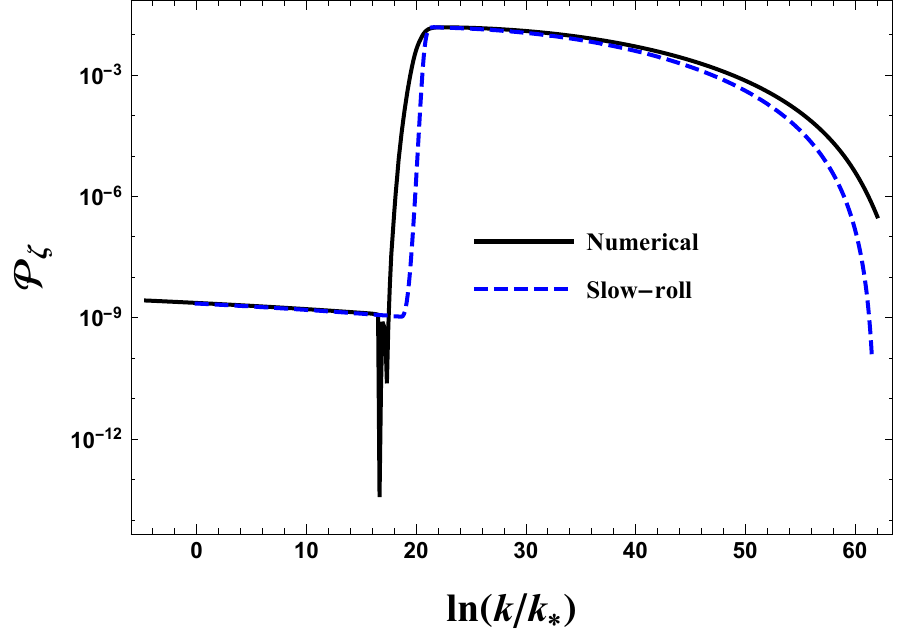}}
\caption{The power spectrum for the polynomial model. The solid line is obtained from the numerical solution and
the dashed line is obtained from the slow-roll approximation \eqref{pwrspec1}. We choose $k_*=0.05$Mpc$^{-1}$.}
\label{fig2}
\end{figure}

\begin{figure}
\centerline{\includegraphics[width=0.6\textwidth]{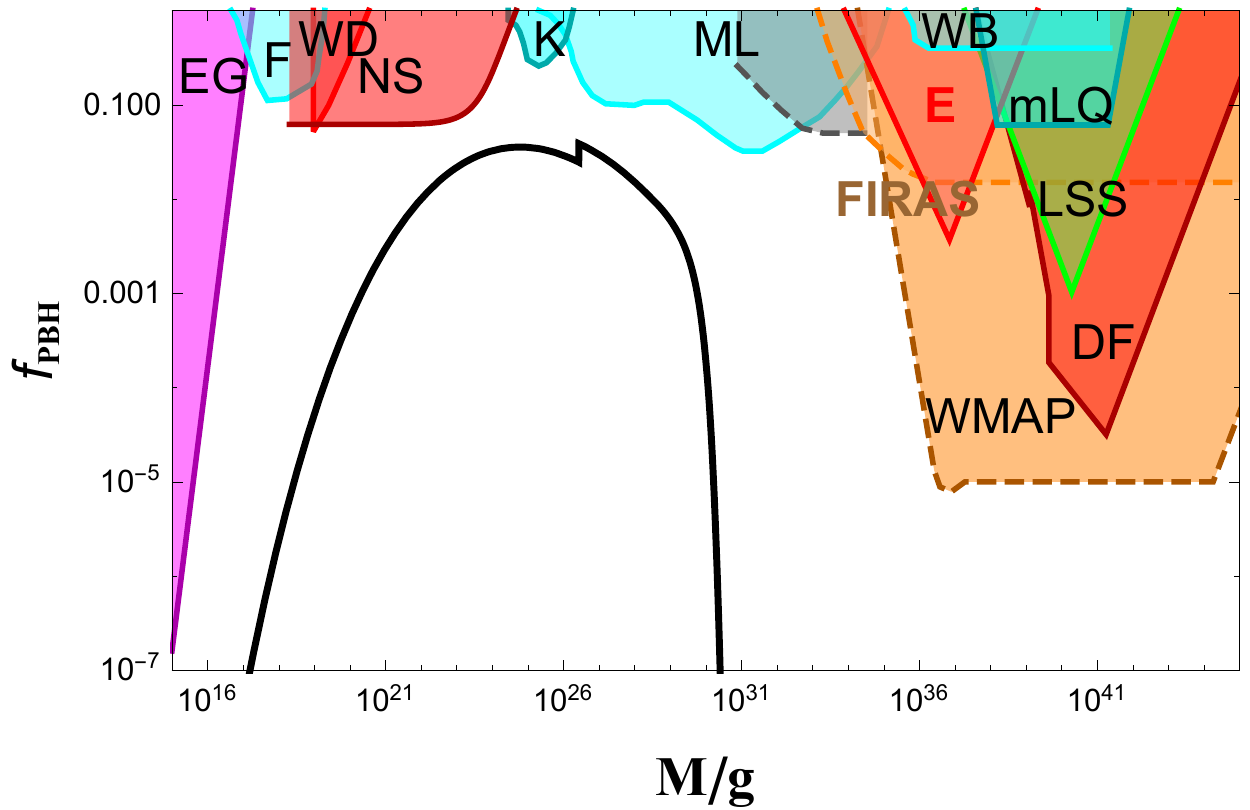}}
\caption{The observational constraints on $f_{\text{PBH}}$ generated by the density perturbation. 
We choose $\delta_c=0.4$, $\Omega_c h^2=0.12$ and $\gamma=3^{-3/2}$ \cite{Carr:1975qj}. 
For the details of the observational constraints, please refer to \cite{Carr:2016drx} and references therein.}
\label{fobsfig}
\end{figure}

To calculate the power spectrum $\mathcal{P}_\zeta=(2\pi^2)^{-1}k^3|v_k/z|^2$, we need to numerically solve
the Mukhanov-Sasaki equation\cite{Mukhanov:1985rz,Sasaki:1986hm},
\begin{equation}
\label{mseq1}
\frac{d^2 v_k}{d\eta^2}+\left(k^2-\frac{1}{z}\frac{d^2z}{d\eta^2}\right)v_k=0,
\end{equation}
where $z=a{\mathcal H}^{-1}d\phi/d\eta=v_k/\zeta$, ${\mathcal H}=a^{-1}da/d\eta$, and the conformal time $d\eta=dt/a$.
The power spectrum from the numerical solution is shown in Fig. \ref{fig2}.
By using the numerical result for the power spectrum, we find that the $\mu$ distortion is $\mu_{\text{ac}}=1.96\times 10^{-8}$ which is
consistent with the observations \cite{Nakama:2017ohe}.
From Fig. \ref{fig2}, we find that the maximum amplitude of the power spectrum is $0.0149$. If we choose $\delta_c=0.12$, we get $\beta=0.027$.
If we choose $\delta_c=0.45$ or $\zeta_c=1.01$ \cite{Germani:2017bcs}, we get $\beta=1.39\times 10^{-16}$.
In Fig. \ref{fobsfig}, we show the current fractional abundance of PBH produced from the power spectrum and
the observational constraints compiled in \cite{Carr:2016drx}.
Therefore, the toy model can produce a sizable amount of PBH dark matter and avoids the problems of large number of $e$-folds and large $\mu$ distortion.

\begin{figure}
\centerline{\includegraphics[width=0.6\textwidth]{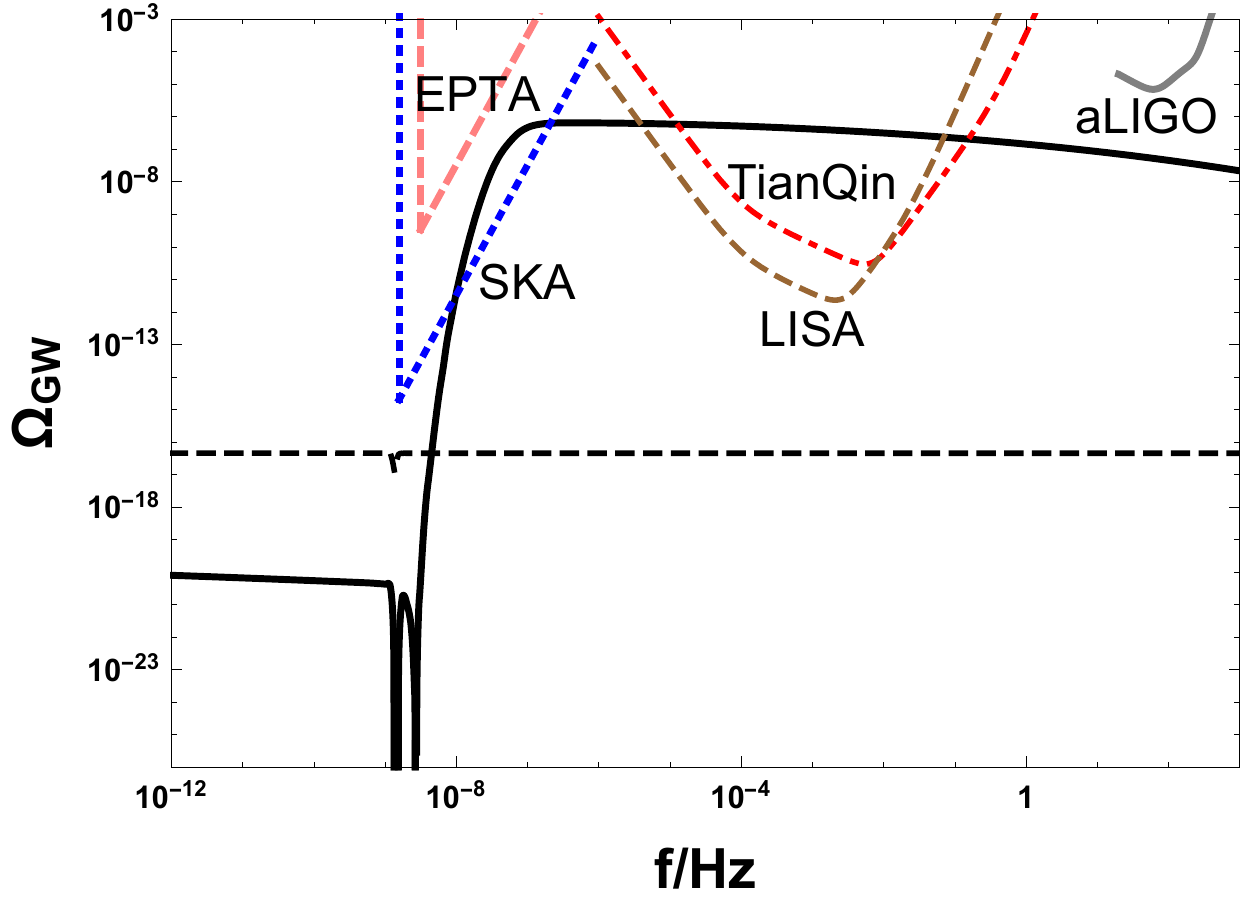}}
\caption{The second order gravitational wave signal generated by the the density perturbations that produce the PBH dark matter.
The black dashed line shows the primary gravitational waves.
The sensitivity curves from different observations are also shown \cite{Sathyaprakash:2009xs,Moore:2014lga,Kuroda:2015owv}.
The pink dashed curve denotes the EPTA limit, the blue dotted curve denotes the SKA  limit, the read dot-dashed curve in the middle denotes the TianQin limit \cite{Luo:2015ght},
the brown dashed curve shows the LISA limit \cite{Audley:2017drz},
and the gray solid curve denotes the aLIGO limit.}
\label{fig4}
\end{figure}

The large density perturbations not only produce the PBH dark matter, but also generate the second order gravitational wave signal \cite{Matarrese:1997ay,Mollerach:2003nq}.
For the small scales ($f=ck/(2\pi)>10^{-12}$ Hz) we are interested in, the modes re-enter the Hubble horizon during the radiation dominated era. During
radiation domination, the Fourier components of the second order tensor perturbations,
for either polarization, satisfy the following equation \cite{Ananda:2006af,Baumann:2007zm,Garcia-Bellido:2017aan,Saito:2009jt,Bugaev:2009zh,Bugaev:2010bb,Alabidi:2012ex}
\begin{equation}
\label{hkeq1}
\frac{d^2 h(\vec{k},\eta)}{d\eta^2}+\frac{2}{\eta}\frac{dh(\vec{k},\eta)}{d\eta}+k^2 h(\vec{k},\eta)=S(\vec{k},\eta),
\end{equation}
where the source term is \footnote{The factor 8 was missed in Eq. (16) in \cite{Ananda:2006af}.}
\begin{equation}
\label{hkeq2}
\begin{split}
S(\vec{k},\eta)=\int \frac{d^3\tilde{k}}{(2\pi)^{3/2}}\tilde{k}^2\left[1-\left(\frac{\vec{k}\cdot\vec{\tilde{k}}}{k\tilde{k}}\right)^2\right]\left\{12\Phi(\vec{k}-\vec{\tilde{k}},\eta)\Phi(\vec{\tilde{k}},\eta)+\right.\\
\left. 8\left[\eta\Phi(\vec{k}-\vec{\tilde{k}},\eta)+\frac{\eta^2}{2}\frac{d\Phi(\vec{k}-\vec{\tilde{k}},\eta)}{d\eta}\right]\frac{d\Phi(\vec{\tilde{k}},\eta)}{d\eta}\right\},
\end{split}
\end{equation}
the conformal time $d\eta=dt/a(t)$, and the Bardeen potential $\Phi=2\zeta/3$ satisfies the equation
\begin{equation}
\label{hkeq3}
\frac{d^2\Phi}{d\eta^2}+\frac{4}{\eta}\frac{d\Phi}{d\eta}+\frac{1}{3}k^2\Phi=0.
\end{equation}
Solving Eq. \eqref{hkeq1} by using the Green function method, we get the current relative energy density of gravitational waves \cite{Baumann:2007zm}
\begin{equation}
\label{hkeq4}
\Omega_{GW}(k,\eta_0)=10\mathcal{P}_\zeta^2 a_{eq},
\end{equation}
where we chose the current scale factor $a(\eta_0)=1$ and $a_{eq}$ is the value of the scale factor at the matter radiation equality.
Take $H_0=67.27$ km/s/Mpc and $\Omega_{m0}=0.3$, we plot the result for $\Omega_{GW}$ in Fig. \ref{fig4}. In Fig. \ref{fig4},
we also show the sensitivity curves \cite{Sathyaprakash:2009xs,Moore:2014lga,Kuroda:2015owv} for the current and future observations from the Pulsar Timing Array (PTA) \cite{Kramer:2013kea,Hobbs:2009yy,McLaughlin:2013ira,Hobbs:2013aka}
including the Square Kilometer Array (SKA) \cite{Moore:2014lga}, the Advanced Laser Interferometer Gravitational-Wave Observatory (aLIGO),
and the TianQin observatory \cite{Luo:2015ght}. From Fig. \ref{fig4}, we see that
the model is consistent with the current PTA observations.
For the half-dome shape of the power spectrum,
the generated second order gravitational waves can be tested by both future PTA and space-borne gravitational wave observations.

\section{Conclusions}

Dependent on the value of $\delta_c$, the maximum value and the enhancement of the power
spectrum needed for the production of appreciable amount of PBH dark matter varies significantly.
To keep the slow-roll conditions and inflation, the decrease in $\epsilon$ cannot happen instantly.
Since $N$ is inversely proportional to $\sqrt{\epsilon}$,
the not-so-fast change of $\epsilon$ will cause the number of $e$-folds spent at $\phi_{\text{infl}}$ to be larger than 60.
This will lead to the enhancement of the power spectrum even at the scale $k=0.1$Mpc$^{-1}$, and cause
the $\mu$ distortion to be large. To overcome these problems for the slow-roll inflation,
an ultra-slow-roll inflation should be reached at $\phi_{\text{infl}}$
because this will cause the number of $e$-folds
spent at $\phi_{\text{infl}}$ to decrease while the enhancement of the power spectrum remains. To reach the
ultra-slow-roll inflation, $V'$ needs to decrease fast before $\phi$ reaches $\phi_{\text{infl}}$ so that it becomes much smaller than
the friction term $3H\dot{\phi}$ at $\phi_{\text{infl}}$,
this condition requires $|\eta|>1$ for a short period of time before $\phi$ reaches $\phi_{\text{infl}}$.
Since $|\eta|>1$ happens for a short period of time, inflation will continue. Because the potential decreases, the inflaton
will not be trapped in the ultra-slow-roll inflation.
Therefore, for the successful production of PBHs from the single field inflation, we
should consider the ultra-slow-roll inflation.
In this paper, we propose the conditions to successfully produce PBHs,
and we use a toy model with the polynomial potential to realize the enhancement of the power spectrum by 7 orders of magnitude.
The number of $e$-folds spent at $\phi_{\text{infl}}$ is $\Delta N=42$ and the maximum amplitude of the power spectrum
is $\mathcal{P}_\zeta=0.0149$, so $\beta=0.027$ if we choose $\delta_c=0.12$ and $\beta=1.39\times 10^{-16}$ if we choose $\delta_c=0.45$.
The model gives $n_s=0.9686$, $r=0.005$, $n_s'=-0.0008$ and $\mathcal{P}_\zeta=2.2\times 10^{-9}$ at $k_*=0.05$Mpc$^{-1}$,
and the $\mu$ distortion is $\mu_{\text{ac}}=1.96\times 10^{-8}$. The power spectrum for the second order gravitational waves generated by the large density perturbations at
small scales is consistent with the current PTA observations, and can be tested by future PTA and space-borne gravitational wave observations.
In conclusion, the model is consistent with current observations.

\begin{acknowledgments}
This research was supported in part by
the Major Program of the National Natural Science Foundation of China under Grant No. 11690021,
and the National Natural Science
Foundation of China under Grant No. 11475065.
\end{acknowledgments}


\providecommand{\href}[2]{#2}\begingroup\raggedright\endgroup

\end{document}